\documentclass[]{spie}  

\usepackage{amsmath,amsfonts,amssymb}
\usepackage{graphicx}
\usepackage[colorlinks=true, allcolors=blue]{hyperref}
\usepackage{siunitx}
\usepackage{braket}
\usepackage{nicefrac}

\title{Generating entangled photon pairs in a parallel crystal geometry}

\author[1]{Alexander Lohrmann}%
\affil[1]{%
 Centre for Quantum Technologies, National University of Singapore, 3 Science Drive 2, S117543\\
}%
\author[1]{Aitor Villar}
\author[1,2]{Alexander Ling}
\affil[2]{Physics Department, National University of Singapore, 2 Science Drive 3, S117542}

\begin{document} 
\maketitle

\begin{abstract}
We present recent findings towards developing brighter entangled photon sources in critically phase matched (CPM) nonlinear crystals. Specifically, we use type-I collinear phase matching at non-degenerate wavelengths in parallel $\beta$-Barium Borate (BBO) crystals to generate pairs of polarization entangled photons for free-space quantum key distribution (QKD). We first review the entangled source configuration and then discuss ways to further improve the source brightness by means of tailoring the pump and collection modes. We present preliminary results that may lead to brighter entangled photon sources that are intrinsically robust to changes in their environment. 
\end{abstract}


\section{INTRODUCTION}
\label{sec:intro}  

Entangled photon pairs are a key component of emerging quantum technologies, such as quantum communication and quantum key distribution \cite{gisin2007quantum,horodecki2009quantum}. Significant effort has been devoted to the development of photon pair sources that exhibit high brightness and high entanglement quality simultaneously. This is in particular important for QKD where the entangled pair rate determines the secure key rate and ultimately limits the communication link distance. Entangled photon sources are most commonly based on spontaneous parametric downconversion (SPDC) in nonlinear crystals, which have yielded a variety of mature source designs with reproducible properties\cite{kwiat1999ultrabright,trojek2008collinear,kim2006phase}. Quantum dots and other 0-dimensional solid state emitters are also rapidly emerging and may be integrated with photonic structures for improved performance in the future\cite{stevenson2006semiconductor,dousse2010ultrabright}.

The current state-of-the-art entangled photon sources are periodically poled quasi-phase matched (QPM) nonlinear crystals which have overtaken traditional critically phase matched sources in terms of brightness and spectral brightness, while maintaining a high state fidelity. The brightest among these are type-0 QPM sources which can generate pair rates exceeding 1 million pairs/second/mW \cite{steinlechner2013phase}. This brightness, however, comes at a cost. QPM relies on temperature tuning and in particular type-0 phase matching is sensitive to small temperature changes. This is not regarded a problem for laboratory applications, but can be a challenge when operating entangled photon sources in aero-space environments.

The dimmer CPM sources, on the other hand, require angle tuning which can be maintained with relative ease, either passively by ensuring high thermo-mechanical stability or actively by using active stabilization methods (e.g., actuators). The brightest CPM sources are based on type-I phase matching (extraordinary pump photon $\rightarrow$ ordinary daughter photons) in a crossed crystal configuration\cite{trojek2008collinear}. In this geometry, the optical axes of two nonlinear crystals are orthogonal to generate the orthogonal polarization components of the entangled state. The drawback is the walk-off induced non-ideal overlap of the emission modes and the highest reported brightness was only \SI{27}{kpairs/\second/\milli\watt} \cite{trojek2008collinear}. To compete with the high brightness QPM sources, it is imperative to improve the brightness of CPM sources further. We have recently shown that by simply rearranging the crystals we can improve the brightness by a factor of 2.4\cite{villar2017experimental}. In this work we first review these results in detail and further show that the new crystal configuration can be improved using elliptical pump beams and free-space detection.

\section{Parallel crystal entangled source}
\label{sec:principle}  

\begin{figure}
\centering
\includegraphics[scale=0.8]{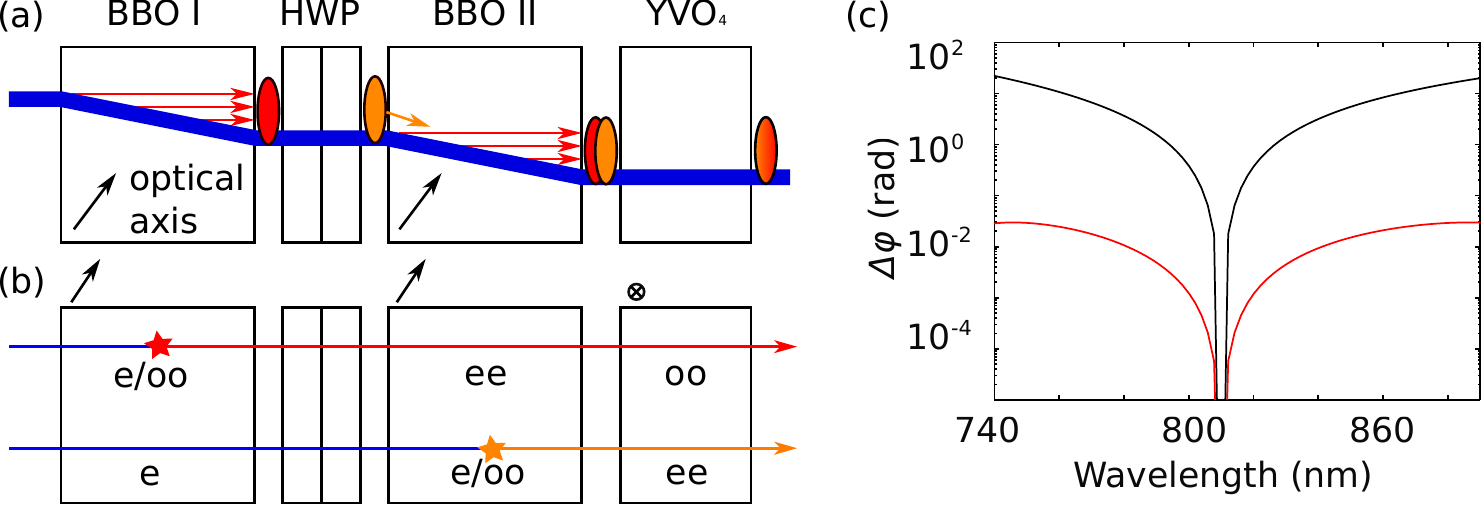}
\vspace{0.3cm}
\caption{(a) Schematic of the SPDC process within the two crystals. The red (orange) ellipse indicates the horizontally (vertically) polarized photons. The black arrows indicate the optical axis orientation. After traversing the temporal compensator, any distiguishability in the time domain is scrambled. (b) Polarization of the individual photons in each crystal (e:~extraordinary, o:~ordinary). The star indicates the downconversion process. The distinguishability arises solely from the pair generated in the first crystal traversing the HWP and the second BBO. The optical axis of the temporal compensation crystal is perpendicular to the propagation direction and \SI{90}{\degree} rotated with respect to the BBO axes (thus: $ee\rightarrow oo$ and vice versa). (c) Phase difference before (black line) and after (red line) temporal compensation (L$_{\text{BBO}}$ = 6~mm, L$_{\text{YVO}}$ = 3.6~mm). A constant offset has been subtracted for clarity.}

\label{fig:concept}
\end{figure}

The concept of a parallel crystal entangled photon source is shown in Fig.~\ref{fig:concept}(a). The vertically polarized pump enters the first crystal and undergoes a spatial walk-off. Along the way, horizontally polarized SPDC pairs are generated. Pump and down converted photons then traverse an achromatic half-wave plate (HWP) that rotates the SPDC polarization state by \SI{90}{\degree}, while leaving the pump polarization unaffected. In the second crystal pump and SPDC photons are both vertically polarized and are affected by the walk-off. Horizontally polarized SPDC photons are again generated along the pump path and exit the crystal spatially overlapping with the emission from the first crystal. The total length of the source lies well within the coherence length of the pump beam.

To produce high entanglement quality, additional phase compensation is required. Without phase compensation, the photon pairs generated in the two crystals are in principle distinguishable by their detector arrival time difference, an effect that is detrimental for entanglement quality. The main reason for this distinguishabilty is that a pair generated in the first crystal propagates through the half-wave plate and the additional BBO, in which the signal and idler components disperse. This dispersion can be negated by an additional birefringent element that induces an additional dispersion based on the photon polarization. In the following, we perform detailed calculations of the phase compensation for the parallel crystal geometry.

The quantum state generated within the crystals is given by,

\begin{equation}
\ket{\Phi} = \frac{1}{\sqrt{2}} \left( \ket{H_sH_i}_2 + e^{i\Delta\varphi} \ket{V_sV_i}_1 \right), 
\label{eq:psi}
\end{equation}

where $\Delta\varphi$ denotes the phase difference and $s$ and $i$ the signal and idler wavelengths, respectively. The phase difference of this state can be directly calculated from the individual photon phases acquired when the photons traverse a non-linear material,

\begin{equation}
\Delta\varphi = \sum_i \varphi_i^H  - \sum_i \varphi_i^V ,
\label{eq:phi}
\end{equation}

where $V$ and $H$ denote the final photon polarization. For a photon passing through the nonlinear material, the phase can be calculated via $\varphi_i =  \nicefrac{2\pi L n}{\lambda}$, where $n$ denotes the refractive index of the material with the length $L$, and $\lambda$ the photon wavelength. The wavelength dependence of $\Delta\varphi$ leads to a mixing of the state shown in Eq.~\ref{eq:psi} which is equivalent to the argument of distinguishability by arrival time difference between signal and idler. To compensate for this, we add an additional birefringent crystal (yttrium orthovanadate, YVO$_4$). 
The total phase difference is then given by,

\begin{align*}
\Delta \varphi = \varphi^ {\text{HWP},V}_{s} +\varphi^ {\text{HWP},V}_{i} +\varphi^{\text{BBO},V}_{s}+ \varphi^{\text{BBO},V}_{i}+\varphi^{ \text{YVO}_4,V}_{s}+\varphi^{ \text{YVO}_4,V}_{i} \\ -(\varphi^{\text{BBO},V}_{p} + \varphi^{\text{HWP},V}_p + \varphi^{ \text{YVO}_4,H}_{s}+ \varphi^{ \text{YVO}_4,H}_{i}).
\end{align*}

Due to the use of a a narrowband pump in this work, all pump terms are neglected. The remaining terms are all determined by the crystal lengths (the half-wave plate consists of magnesium floride and quartz) and the refractive index of the material for the respective photon polarization and wavelength. The individual photon polarization in each crystal is shown in Fig.~\ref{fig:concept}(b). By solving the equation $\nicefrac{\partial(\Delta\varphi)}{\partial \lambda} = 0$ for the temporal compensator length, the right YVO$_4$ length can be selected. The constant phase difference can then be adjusted for example by tilting any nonlinear crystal to any value between $C$ and $2\pi+C$ ($C$ denotes a constant phase offset) to generate one of the two Bell states, $\Phi^+$ or $\Phi^-$. It should be noted that emission processes from near the entry (exit) face of the first crystal are overlapping with those originating from the entry (exit) face of the second crystal. This overlap ensures a constant phase difference across the whole walk-off pattern and prevents mixing of the maximally entangled Bell state for collinear emission. An example for the phase difference before and after temporal compensation is shown in Fig.~\ref{fig:concept}(c). The ideal compensator lengths scales linearly with the BBO lengths with $L_\text{YVO} = 0.52L_\text{BBO} + C_\text{HWP}$, where a constant offset is introduced by the phase acquired in the HWP.


We have demonstrated the successful operation of this source recently\cite{villar2017experimental} using 5~mm BBOs and a temporal compensator length of 3.12~mm. The pair rate of this source was \SI{65}{\kilo pairs/\second/\milli\watt} with signal and idler efficiencies of 0.27 and 0.22, respectively. The efficiencies were mainly limited by the detection efficiencies of the single photon detectors and the transmission losses of the downstream optics. From the efficiencies we can estimate a total pair generation rate of approximately \SI{1.1}{\mega pairs/\second/\milli\watt}. We estimated a state fidelity towards the maximally entangled Bell state, $\Phi^-$, of greater than 99\%.

As the full pump power can be used in each crystal, a factor of 2 increase in the collected pair rate is expected when compared to configurations that utilize crossed crystals\cite{trojek2008collinear}. Our results indicate an increased brightness of a factor of 2.4. The additional gain was achieved by elliptical focusing of the pump as explained in the following section.

\section{Pump considerations}
\label{sec:elliptical}

\begin{figure}
\centering
\includegraphics[scale=0.8]{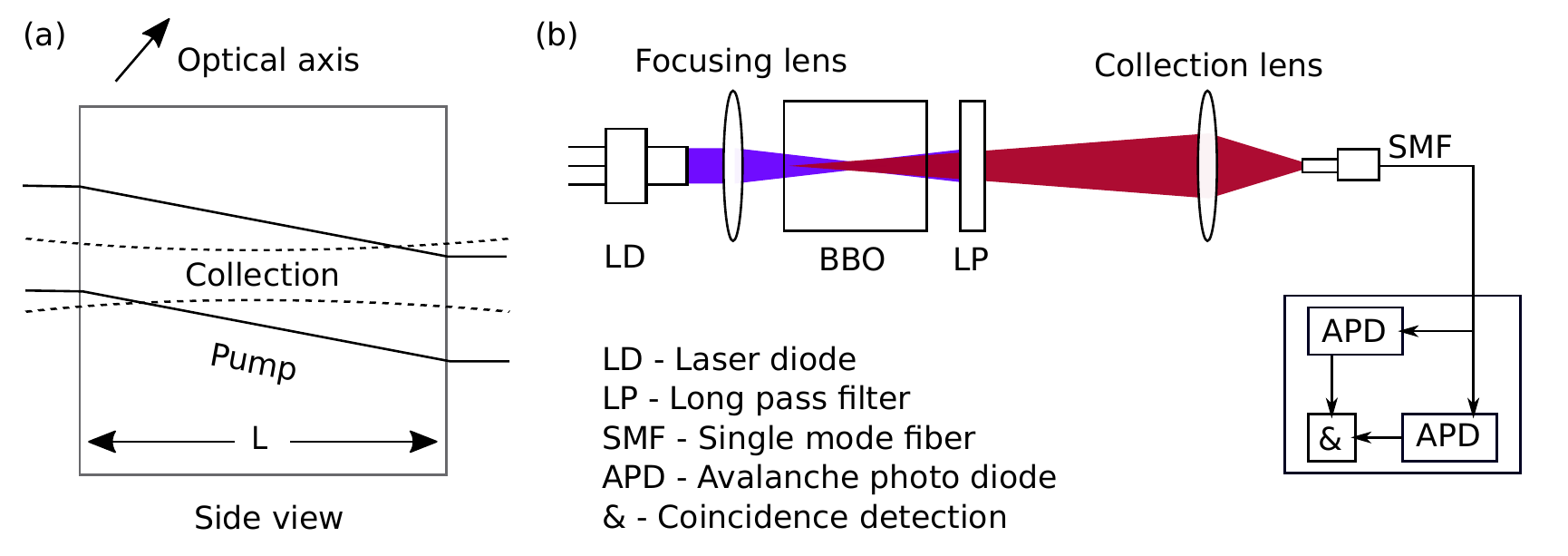}
\vspace{0.3cm}
\caption{(a) Sketch of the overlap of pump and collection in the presence of walk-off. (b) Schematic of the experimental setup to investigate the impact of beam ellipticity on SPDC brightness.}
\label{fig:beamProfiles}
\end{figure}

To optimize the brightness of an SPDC source, it is imperative to investigate the ideal pump and collection parameters. In most common SPDC sources\cite{kwiat1999ultrabright,kim2006phase,trojek2008collinear,steinlechner2013phase}, circular symmetric pump and collection conditions are used as proposed in early studies on parametric processes\cite{boyd1968parametric}. Tightly focusing both waists (pump: $\omega_p$, collection: $\omega_{s/i}$ ) increases the total collected brightness of the source\cite{boyd1968parametric}. However, the walk-off in critically phase-matched crystals limits the minimal waist size in walk-off direction, as it reduces the spatial overlap of emission from the entry and exit face, as indicated in Fig.~\ref{fig:beamProfiles}(a). 

This limitation is not present for the waist size perpendicular to the walk-off direction. Therefore, shaping the pump to an elliptical beam profile can be beneficial in parametric processes. In particular, it was shown that elliptical focusing can improve brightness in second harmonic generation\cite{freegarde1997general} and parametric gain\cite{kuizenga1972optimum}. This is even more interesting when considering that most SPDC sources are pumped by laser diodes which exhibit intrinsically elliptical beam profiles. This approach has been overlooked for entangled photon sources, because in the crossed crystal geometry\cite{kwiat1999ultrabright,trojek2008collinear} an elliptical pump would cause an imbalance in the H and V components, diminishing the improvement. In the present work, on the other hand, the optical axes are parallel allowing us to make use of an elliptical pump shape.

\begin{figure}
\centering
\includegraphics[scale=0.8]{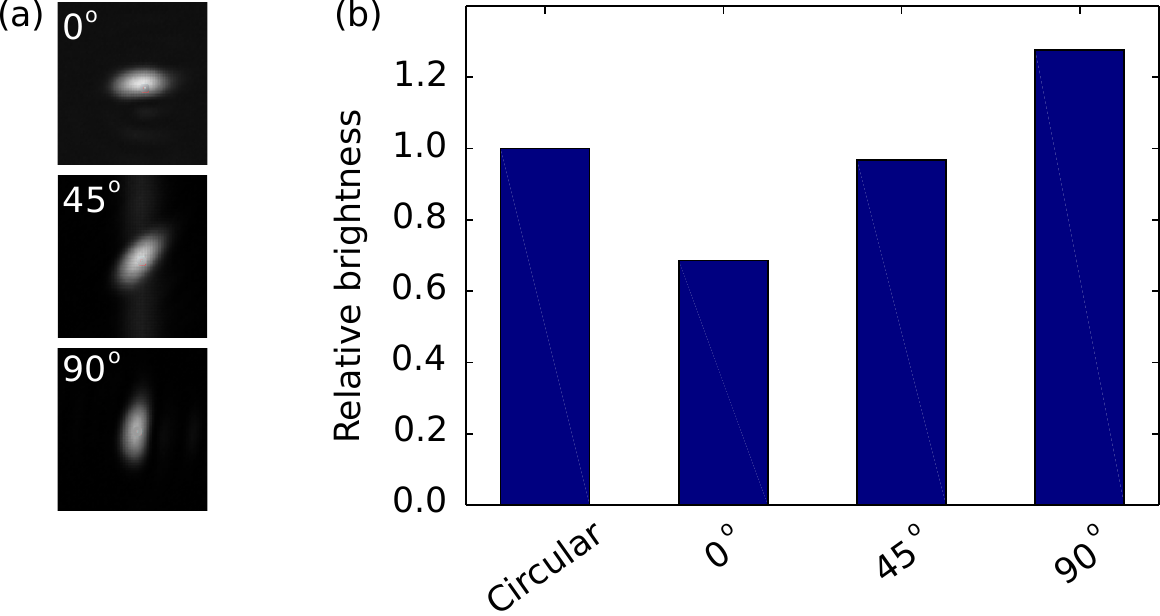}
\vspace{0.3cm}
\caption{(a) Three different pump orientations with the same aspect ratio (2:1). The collection conditions remained the same. (b) Relative brightness of the three different orientations.}
\label{fig:beamProfiles_res}
\end{figure}

As a proof of concept, we compared the ideal (single-mode fiber filtered) circular symmetric pump for a single BBO, to an elliptical pump as emitted from the laser diode. The empirically found optimal circular symmetric waist size is approximately \SI{100}{\micro\metre} with an optimal collection waist of \SI{45}{\micro\metre}. Without changing the collection conditions, we replace the circular symmetric beam with the elliptical beam with an aspect ratio of 2:1 and a major axis waist kept at the ideal circular symmetric value of \SI{100}{\micro\metre}. 

The setup is shown schematically in Fig.~\ref{fig:beamProfiles} (b). The output from a narrowband laser diode (aspect ratio 2:1) was focused into a BBO crystal using a single lens. After filtering out the pump, the SPDC photons were collected into a single mode fiber using a single collection lens. The collection mode was therefore circular-symmetric. Signal and idler photons were split and detected separately. The pump orientation could be rotated around the beam axis. We measured the relative change in brightness compared to the circular symmetric pump for three different ellipse orientations, as displayed in Fig.~\ref{fig:beamProfiles_res}(a). The pump polarization was extraordinary in all cases.

Figure~\ref{fig:beamProfiles_res}(b) shows the relative brightness for the different orientations. It is apparent that the elliptical pump, when oriented in walk-off direction, leads to an increase in the total brightness of approximately 1.3. This is in good agreement with the additional brightness gain in the parallel crystal entangled source\cite{villar2017experimental}. Shrinking the waist in walk-off direction, on the other hand, reduces the total brightness. Work is currently on-going to investigate the ideal pump aspect ratios in critically phase-matched entangled photon sources. It should be noted that further improvement can be expected when the minor axis is focused beyond the values presented here.

\begin{figure}[t]
\centering
\includegraphics[scale=0.8]{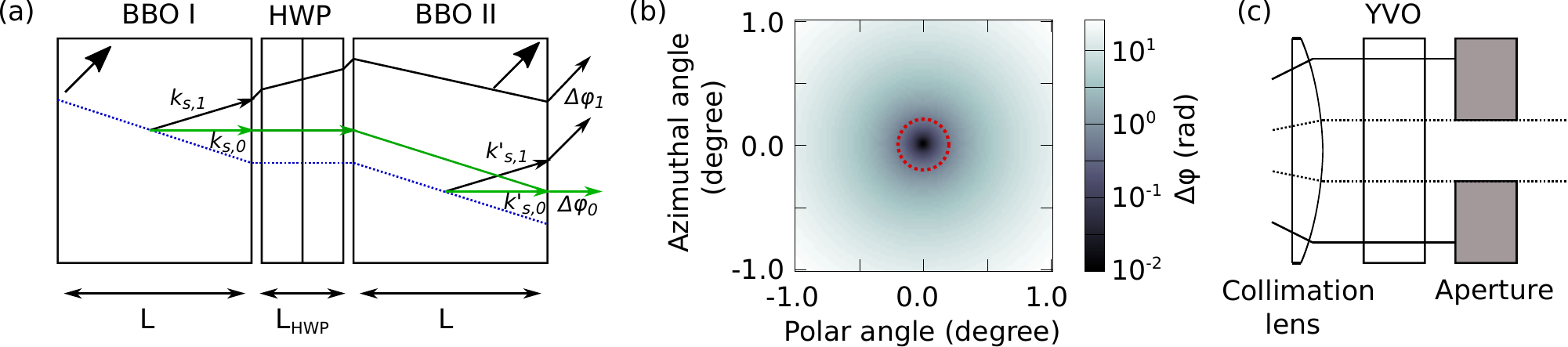}
\vspace{0.3cm}
\caption{(a) Sketch of the SPDC pair (only signal shown, $k_{s, 0/1}$) generated in the first BBO traversing the HWP and the second BBO. As more non-linear material is traversed for non-collinear emission, pairs collected with a greater opening angle (black) exhibit a larger phase difference than the collinear emission (green). Laser path indicated in blue (dotted line). (b) Calculations of the total phase difference for azimuthal and polar free-space propagation angles for a source based on 6~mm BBO crystals. The red circle indicates a region with approximately constant phase. (c) After collimation and wavelength compensation, an aperture can be placed in the beam to experimentally restrict the opening angle to the region indicated in (b).}
\label{fig:multimode}
\end{figure}

\section{Multimode collection}
\label{sec:phase}

\begin{figure}[b]
\centering
\includegraphics[scale=0.8]{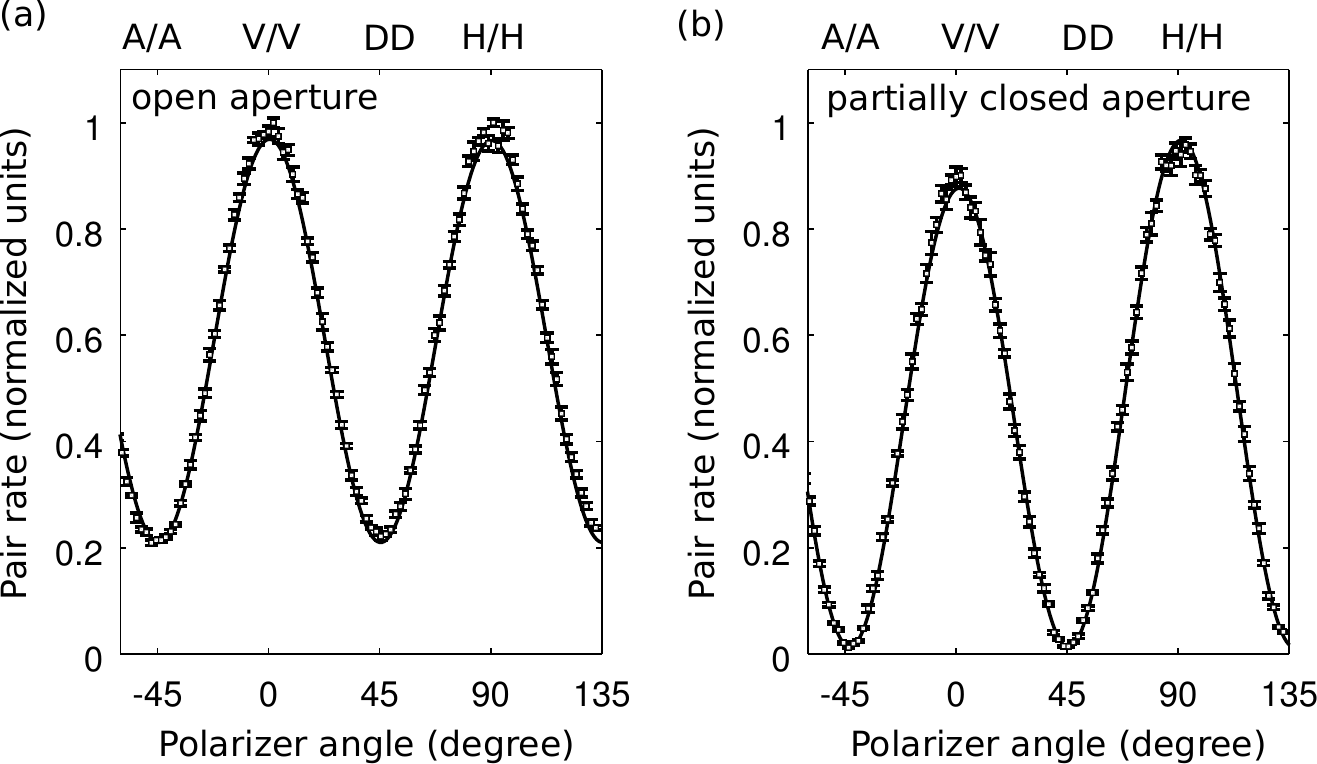}
\vspace{0.3cm}
\caption{Polarization sweeps of a single polarizer in the SPDC path with (a) a fully opened, and (b) a partially closed aperture. The maxima and minima indicate the projection of signal and idler into the linear bases (H/V and A/D). The measurement shown in (b) is close to the ideal maximally entangled Bell state, $\Phi^-$.}
\label{fig:multimode_res}
\end{figure}

In the first demonstration of the parallel source\cite{villar2017experimental}, we used single mode fiber collection which negates all angle-dependent phase effects. Due to the spatiotemporal self-compensation and the overlapping of the emission modes from the two crystals, collection with a free-space detector instead of a single mode fiber can be attempted. This allows in principle collection of higher pair rates. However, pairs originating from the first crystal traverse the HWP and the second crystal under an angle and pick up an angle-dependent phase. This introduces an additional phase difference, as indicated in Fig.~\ref{fig:multimode}(a) with $\Delta\varphi_1 > \Delta\varphi_0$. This effect and its compensation have previously been demonstrated in non-collinear emission in thin crystals\cite{kwiat1999ultrabright,altepeter2005phase}. While in the early works only the angular dependence in the phase matching plane was considered, it is clear that it also exists for the orthogonal plane\cite{hegazy2017relative}. Moreover, previous reports have generally focused on short crystals ($\leq 1$~mm) for which angle effects are minor.

The angle-dependent phase difference can be calculated directly from Eq.~\ref{eq:phi} where the traversed crystal lengths and refractive indices depend on the emission angles. We evaluated the total phase difference for the azimuthal and polar free-space propagation angles (see Fig.~\ref{fig:multimode}(b)). A flat region can be identified (dashed red circle) over which the phase difference remains approximately constant. To collect the maximally entangled state, we restrict the angular spread of collected photons to the flat region depicted in Fig.~\ref{fig:multimode}(b).

We test this hypothesis by collecting the SPDC emission of the entangled photon source using multi-mode fibers (core diameter: \SI{62.5}{\micro\metre}, NA = 0.22) which is in our case equivalent to using a free-space detector. To restrict the emission angle an aperture is placed in the SPDC beam and the source brightness and polarization curve with and without the aperture are measured. A single polarizer after the temporal compensator before splitting signal and idler photons is used to assess the polarization state. The results of the measurements are shown in Fig.~\ref{fig:multimode_res}. Without the aperture, a source brightness of \SI{312}{\kilo pairs/\second/\milli \watt} was measured with a visibility, $V = \frac{B_{max}-B_{min}}{B_{max}+B_{min}}$ ($B$ denotes the measured pair rate), of $V=0.67$. With the partially closed aperture, the pair rate reduced to  \SI{78}{\kilo pairs/\second/\milli \watt}. The visibility, on the other hand, reached values of 98\%.

Assuming that the source produces only $\ket{HH}$ and $\ket{VV}$ components and a maximally entangled target state, $\Phi^-$, the visibility of the polarization sweep can be used to assess the entanglement quality. The visibility is approximately equivalent to that of the A/D basis in a two polarizer measurement.

It is important to note that this angle filtering is not possible to this extent in previous designs with thick crossed crystals where a spatial phase difference across the emission pattern exists. Work is currently ongoing to investigate the maximally achievable brightness and visibility for free-space detection and to experimentally reproduce the calculated phase maps. Moreover, we are investigating ways to compensate the angle-dependent phase difference using additional birefringent crystals.

\section{Conclusion}
We have demonstrated an experimental realization of an entangled photon source using parallel nonlinear crystals. In essence, the source is a simplified version of the double pass geometry \cite{steinlechner2013phase}, but avoids most of its drawbacks (such as mirror reflectivity, walk-off compensation, polarization effects from dichroic beam splitters). The demonstrated total pair rates per unit of crystal length are still lower than those in periodically poled crystals, but some improvement can be expected. The use of a high eccentricity pump may lead to further improvement of the total collected rates in case of single mode fiber collection. In the case of free-space collection, compensating the angular phase difference by additional non-linear crystals may boost the brightness by an order of magnitude. Moreover, longer crystals can be used to further improve the brightness. Future work will enable the brightness of this parallel crystal design to catch up with QPM-based sources.

Lastly, we note that the produced pair rates are already in a regime for which local detection of one photon will saturate commonly used Si-APDs. Therefore many quantum communication protocols are limited by the efficiency and dead time of single photon detectors. In this case, a passively stabilized critically phase matched system may prove advantageous over quasi phase-matched systems.

\section{Acknowledgements}
This material is based upon work supported by the Air Force Office of Scientific Research under award number FA2386-17-1-4008. This program is supported by the National Research Foundation, Prime Minister’s Office of Singapore, and the Ministry of Education, Singapore.


\bibliography{bibliography} 

\begin{thebibliography}{10}

\bibitem{gisin2007quantum}
Gisin, N. and Thew, R., ``Quantum communication,'' {\em Nature photonics}~{\bf
  1}(3),  165 (2007).

\bibitem{horodecki2009quantum}
Horodecki, R., Horodecki, P., Horodecki, M., and Horodecki, K., ``Quantum
  entanglement,'' {\em Reviews of modern physics}~{\bf 81}(2),  865 (2009).

\bibitem{kwiat1999ultrabright}
Kwiat, P.~G., Waks, E., White, A.~G., Appelbaum, I., and Eberhard, P.~H.,
  ``Ultrabright source of polarization-entangled photons,'' {\em Physical
  Review A}~{\bf 60}(2),  R773 (1999).

\bibitem{trojek2008collinear}
Trojek, P. and Weinfurter, H., ``Collinear source of polarization-entangled
  photon pairs at nondegenerate wavelengths,'' {\em Applied Physics
  Letters}~{\bf 92}(21),  211103 (2008).

\bibitem{kim2006phase}
Kim, T., Fiorentino, M., and Wong, F.~N., ``Phase-stable source of
  polarization-entangled photons using a polarization sagnac interferometer,''
  {\em Physical Review A}~{\bf 73}(1),  012316 (2006).

\bibitem{stevenson2006semiconductor}
Stevenson, R.~M., Young, R.~J., Atkinson, P., Cooper, K., Ritchie, D.~A., and
  Shields, A.~J., ``A semiconductor source of triggered entangled photon
  pairs,'' {\em Nature}~{\bf 439}(7073),  179 (2006).

\bibitem{dousse2010ultrabright}
Dousse, A., Suffczy{\'n}ski, J., Beveratos, A., Krebs, O., Lema{\^\i}tre, A.,
  Sagnes, I., Bloch, J., Voisin, P., and Senellart, P., ``Ultrabright source of
  entangled photon pairs,'' {\em Nature}~{\bf 466}(7303),  217 (2010).

\bibitem{steinlechner2013phase}
Steinlechner, F., Ramelow, S., Jofre, M., Gilaberte, M., Jennewein, T., Torres,
  J.~P., Mitchell, M.~W., and Pruneri, V., ``Phase-stable source of
  polarization-entangled photons in a linear double-pass configuration,'' {\em
  Optics express}~{\bf 21}(10),  11943--11951 (2013).

\bibitem{villar2017experimental}
Villar, A., Lohrmann, A., and Ling, A., ``Experimental entangled photon pair
  generation using parallel crystals,'' {\em arXiv preprint arXiv:1711.01045}
  (2017).

\bibitem{boyd1968parametric}
Boyd, G. and Kleinman, D., ``Parametric interaction of focused gaussian light
  beams,'' {\em Journal of Applied Physics}~{\bf 39}(8),  3597--3639 (1968).

\bibitem{freegarde1997general}
Freegarde, T., Coutts, J., Walz, J., Leibfried, D., and H{\"a}nsch, T.,
  ``General analysis of type {I} second-harmonic generation with elliptical
  gaussian beams,'' {\em JOSA B}~{\bf 14}(8),  2010--2016 (1997).

\bibitem{kuizenga1972optimum}
Kuizenga, D., ``Optimum focusing conditions for parametric gain in crystals
  with double refraction,'' {\em Applied Physics Letters}~{\bf 21}(12),
  570--572 (1972).

\bibitem{altepeter2005phase}
Altepeter, J.~B., Jeffrey, E.~R., and Kwiat, P.~G., ``Phase-compensated
  ultra-bright source of entangled photons,'' {\em Optics Express}~{\bf
  13}(22),  8951--8959 (2005).

\bibitem{hegazy2017relative}
Hegazy, S.~F., Badr, Y.~A., and Obayya, S.~S., ``Relative-phase and time-delay
  maps all over the emission cone of hyperentangled photon source,'' {\em
  Optical Engineering}~{\bf 56}(2),  026114 (2017).

\end{thebibliography}
\bibliographystyle{spiebib} 

\end{document}